\documentclass[12pt]{revtex4}
\usepackage{amsmath,amssymb}
\input epsf
\usepackage{verbatim}
\usepackage{graphicx}
\usepackage{bm}
\usepackage{natbib}
\def\pmb#1{\setbox0=\hbox{#1}
\kern-.025em\copy0\kern-\wd0 \kern-.05em\copy0\kern-\wd0
\kern-.025em\raise.0433em\box0}

\newcommand{\C}{\mathbb C}

\newcommand{\beq}{\begin{equation}}
\newcommand{\eeq}{\end{equation}}
\newcommand{\ba}{\begin{eqnarray}}
\newcommand{\ea}{\end{eqnarray}}

\usepackage[english]{babel}
\begin{document}

\title[Elastodynamic cloaking and field enhancement for soft spheres]{Elastodynamic cloaking and field enhancement for soft spheres
}
\author{Andre Diatta$^{(1)}$
 and  Sebastien Guenneau$^{(1)}$}

\address{ $^{(1)}$Aix$-$Marseille Univ., CNRS, Centrale Marseille, Institut Fresnel, UMR CNRS 7249, 13013 Marseille, France}

\begin{abstract}
We propose a spherical cloak described by a non-singular asymmetric elasticity tensor ${\C}$ depending upon a small parameter $\eta,$  that defines the softness of a region one would like to conceal from elastodynamic waves. 
By varying $\eta$, we generate a class of soft spheres dressed by elastodynamic cloaks, which are shown to considerably reduce the soft spheres' scattering. Importantly, such cloaks also provide some wave protection except for a countable set of frequencies, for which some large elastic field enhancement  can be observed within the soft spheres. 
Through an investigation of trapped modes in elasticity,  we supply a good approximation of such Mie-type resonances by some transcendental equation. Our results, unlike previous studies that focused merely on invisibility aspects, shed light on potential pitfalls of elastodynamic cloaks for earthquake protection designed via  geometric transforms: A seismic cloak needs to be designed in such a way that its inner resonances differ from eigenfrequencies of the building one wishes to protect. 
In order to circumvent this downfall  of field enhancement inside the cloaked area, we introduce a novel generation of cloaks, named here, mixed cloaks. Such mixed cloaks consist of a shell that detours incoming waves, hence creating an invisibility region, and of a Perfectly Matched Layer (PML, located at the inner boundary of the cloaks) that absorbs residual wave energy in such a way that aforementioned resonances in the soft sphere are strongly attenuated.

Designs of mixed cloaks with non-singular elasticity tensor combined with an inner PML and non-vanishing density bring seismic cloaks one step closer to a  practical implementation. Note in passing that the concept of mixed cloaks also applies in the case of singular cloaks and can be translated in other
wave areas for similar purpose (i.e. to smear down inner resonances within the invisibility region). 
\end{abstract}
\maketitle
\section{Introduction}
Ten years ago, two groups of physicists unveiled theoretical paths towards the design of invisibility cloaks for electromagnetic waves, by making geometric transforms in the Maxwell's \cite{pendry} and Helmholtz's \cite{leonhardt} equations. Such transforms conceal a region of space (e.g. by blowing up a point to a circle, or a sphere). This prompted mathematicians to investigate coordinate changes in other wave equations, notably the Navier equations for elastic waves  
\cite{milton2006}. The main difference between Maxwell's and Navier's equations is that the former retain their form under coordinate changes, unlike the latter. Notably, in 
 \cite{brunapl} the design of a cylindrical cloak for in-plane elastic waves with an asymmetric elasticity tensor resulting from a radially symmetric geometric transform was investigated numerically as a modified version of the Willis' type
 \cite{willis1981}  transformed equations derived in 
 \cite{milton2006}. The seemingly irreconcilable structures of transformed Navier equations studied in 
\cite{milton2006,  brunapl}  were later encompassed within a more general elasticity framework
  \cite{norris2011}. In parallel to the developments of these theoretical and numerical works, engineering science has gained a prominent position in the metamaterials' community, as more and more groups worldwide get to work on the design, fabrication and characterization of so-called acoustic metamaterials \cite{21}. Notably, water wave
\cite{prl2008} and acoustic 
\cite{cummer-schurigt2007,sanchez2007,chen2007,norris2008,cummer-pendry-prl-mie} cloaks have been investigated. Acoustic metamaterials for lensing \cite{20} and other transformation based imaging systems have given rise to a flourishing literature \cite{21}. Regarding bulk elastic waves, some theoretical 
\cite{farhat2009,farhat2012,alu} and experimental 
\cite{stenger2012} progress has been made in the control of flexural elastic waves in thin plates. In that case, the transformed governing equations (e.g. Kirchhoff) have a simpler structure which helps engineer structured cloaks.

In the present article, we investigate spherical cloaks for solid elastic waves using a radially symmetric  linear one-to-one geometric transform that depends upon a positive parameter $\eta$ no greater than 1. See Figure \ref{3d_cloak_sphere_elastic1}-(A) and  Figure \ref{3d_cloak_sphere_elastic1}-(B) for schematic illustrations.   The resulting cloak models an elastodynamic medium, with nonsingular elasticity tensor and finite nowhere zero scalar mass density, that  can be viewed as a generalization of the work \cite{diatta-guenneauAPL2014}. The latter corresponds to the case when the parameter $\eta$ is zero.
However, in \cite{diatta-guenneauAPL2014} we assumed some stress-free boundary conditions on the inner boundary of the cloak so that one cannot observe the phenomenon of wave protection within that paper. We note that, for small values of the parameter $\eta$, the cloaked region is a soft sphere, which is hence sensitive to any field that may possibly bypass the cloak and penetrate it.
Let us stress that an object placed within the cloak is completely invisible to an external  observer, hence it is different from shrinking devices studied by some authors \cite{shrinking}.

We discuss their underlying mechanism and illustrate
the theory using a finite element approach which is adequate to solve the Navier
equations in transformed anisotropic heterogeneous media
with asymmetric elasticity tensors.

The numerical exploration unveils a countable set of resonant eigenfrequencies, for which the elastic field is  enhanced (trapped elastic modes) within the soft spheres, thereby making protection, say, against seismic waves, not a trivial consequence of cloaking.

 However,  the cloaks provide some wave protection outside such a countable set of  Mie-type resonances that are well approximated by some transcendental equation (\ref{resonances}).
We propose a new generation of cloaks, the  mixed cloaks, as a way to circumvent this downfall  of field enhancement inside the cloaked area at specific resonances. Figure \ref{3d_cloak_sphere_elastic1}-(C) gives a schematic picture of the geometric construction of mixed cloaks. These consist of two concentric spherical shells sharing a boundary. The first (outer) shell acts as an ordinary elastodynamic cloak, that is, it detours incoming waves so that, the enclosed region becomes an invisibility region. The second (inner) shell is a Perfectly Matched Layer that absorbs residual wave energy in such a way that aforementioned resonances in the (enclosed) soft sphere are strongly attenuated. Let us emphasize that here we
use PML not as a computational tool to model unbounded domains as in section IV.A, but rather as a mean to attenuate (without reflection) a wave within the cloak. In practice, one could achieve mixed cloaks through a homogenization algorithm for both the PML and the cloak. The fact that these two shells are described by asymmetric elasticity tensors means that classical homogenization would fail, but there exists some subtle way to approach the ideal PML and cloak parameters by layers of symmetric, homogeneous and isotropic elastic media \cite{Diatta-Guenneau-PML2016}.

The propagation of
elastic waves is governed by the Navier equations. Assuming
time harmonic $\exp(-i\omega t)$ 
dependence, with $\omega$ as the
angular wave frequency and $t$ the time variable, allows us to work directly in the spectral domain.
Such dependence is assumed henceforth and suppressed leading, in spherical coordinates, to
\begin{eqnarray}
\begin{array}{lll}
- \sqrt{-1}~ \omega  ~ p_r&=\frac{1}{r^2}\frac{\partial }{\partial r}\left( r^2 \sigma_{rr} \right)+ \frac{1}{r}\left(   \frac{\cos\theta}{\sin\theta}
+ \frac{\partial}{\partial \theta} \right)\sigma_{\theta r}\\
& -\frac{1}{r} \sigma_{\theta \theta}+  \frac{1}{r\sin\theta} \frac{\partial}{\partial \phi}  \sigma_{\phi r} -  \frac{1}{r}\sigma_{\phi\phi} \\
- \sqrt{-1}~ \omega  ~ p_\theta &=\frac{1}{r^2}\frac{\partial }{\partial r}\left( r^2 \sigma_{r\theta} \right)+ \frac{1}{r}\left(   \frac{\cos\theta}{\sin\theta}
+ \frac{\partial}{\partial \theta} \right)\sigma_{\theta\theta}
\\
& + \frac{1}{r} \sigma_{\theta r}+  \frac{1}{r\sin\theta} \frac{\partial}{\partial \phi}  \sigma_{\phi \theta} -  \frac{\cos\theta}{r\sin\theta}\sigma_{\phi\phi} \\
- \sqrt{-1}~ \omega  ~ p_\phi &=\frac{1}{r^2}\frac{\partial }{\partial r}\left( r^2 \sigma_{r\phi} \right)+ \frac{1}{r}\left(   \frac{\cos\theta}{\sin\theta}
+ \frac{\partial}{\partial \theta} \right)\sigma_{\theta\phi}
\\
& +\frac{1}{r} \sigma_{\phi r}+ \frac{\cos\theta}{r\sin\theta} \frac{\partial}{\partial \phi}  \sigma_{\phi \theta} + 
 \frac{1}{r\sin\theta}  \sigma_{\phi\phi}
\end{array}
\label{navier1}
\end{eqnarray}
where the quantity of motion ${\bf p} = (p_j)$ and the stress tensor ${\boldsymbol{\sigma}} =(\sigma_{ij})$ are related to the
displacement field ${\bf u}=(u_j)$ via
\begin{eqnarray}
\begin{array}{ll}
\sigma_{ij}=C_{ijkl} ~\left(\nabla{\mathbf u}\right)_{kl} \; , \; p_j=- \sqrt{-1}~\omega~\rho~ u_j \; ,
\label{navier2}
\end{array}
\end{eqnarray}
with $i,j,k,l=r,\theta,\phi\;  $ , where $\C:={(C_{ijkl})}$ is the rank-four elasticity tensor,
$\nabla{\mathbf u}$ is the deformation tensor, see (\ref{gradu}), and  $\rho$ the scalar density of the elastic medium.

From now on, we consider Equation (\ref{navier1}) in an isotropic homogeneous elastic medium with Lam\'e parameters $\lambda$ and $\mu$ , so that $\C$ has the following 21 non-zero coefficients:
$
C_{iiii}=  \lambda+2\mu
$
and if 
$i\neq j,  \;\; C_{iijj}= \lambda\; , \;  C_{ijij}=  C_{ijji}=\mu.
$  See Appendix A. 

Following  \cite{norris2011}, let us consider the coordinate change ${\bf x}=(r,\theta,\phi)\longmapsto {\bf x}'=(r',\theta',\phi')$  and assume that the resulting transformed displacement ${\bf u'} = {\bf u'}(r',\theta',\phi')$ is linearly related to ${\bf u}$ as 
${\bf u}={\bf A}^T{\bf u'},$ with ${\bf A}$ a matrix field in general. Below, we will let {\bf S} and {\bf D} denote some third order tensors possibly encompassing an $\omega$ dependence.  
The above coordinate change leads to
a transformed equation
\begin{eqnarray}
\begin{array}{lll}
- \sqrt{-1}~ \omega  ~ p_{r'}'&=\frac{1}{r'^2}\frac{\partial }{\partial r'}\left( r'^2 \sigma_{r'r'}' \right)+ \frac{1}{r'}\left(   \frac{\cos\theta'}{\sin\theta'}
+ \frac{\partial}{\partial \theta'} \right)\sigma_{\theta' r'}'\\
& -\frac{1}{r'} \sigma_{\theta' \theta'}'+  \frac{1}{r'\sin\theta'} \frac{\partial}{\partial \phi'}  \sigma_{\phi' r'}' -  \frac{1}{r'}\sigma_{\phi'\phi'}' \\
- \sqrt{-1}~ \omega  ~ p_{\theta'}' &=\frac{1}{r'^2}\frac{\partial }{\partial r'}\left( r'^2 \sigma_{r'\theta'}' \right)+ \frac{1}{r'}\left(   \frac{\cos\theta'}{\sin\theta'}
+ \frac{\partial}{\partial \theta'} \right)\sigma_{\theta'\theta'}'
\\
& + \frac{1}{r'} \sigma_{\theta' r'}'+  \frac{1}{r'\sin\theta'} \frac{\partial}{\partial \phi'}  \sigma_{\phi' \theta'} '-  \frac{\cos\theta'}{r'\sin\theta'}\sigma_{\phi'\phi'}' \\
- \sqrt{-1}~ \omega  ~ p_{\phi'}' &=\frac{1}{r'^2}\frac{\partial }{\partial r'}\left( r'^2 \sigma_{r'\phi'}' \right)+ \frac{1}{r'}\left(   \frac{\cos\theta'}{\sin\theta'}
+ \frac{\partial}{\partial \theta'} \right)\sigma_{\theta'\phi'}'
\\
& +\frac{1}{r'} \sigma_{\phi' r'}'+ \frac{\cos\theta'}{r'\sin\theta'} \frac{\partial}{\partial \phi'}  \sigma_{\phi' \theta'}' + 
 \frac{1}{r'\sin\theta'}  \sigma_{\phi'\phi'}',
\end{array}
\label{navier3}
\end{eqnarray}
whereby the 
transformed $ {\bf p'} = (p_{j'}')$, stress  tensor ${\boldsymbol{\sigma}'} =(\sigma_{i'j'}')$, rank-four elasticity tensor $\C'=(C_{i'j'k'l'}')$ and  density $\rho'=(\rho_{i'j'}'),$  are now related to the 
transformed  displacement ${\bf u}'={\bf u'}(r',\theta',\phi')$,  by 
\begin{eqnarray}
\begin{array}{ll}
\sigma_{i'j'}'=C_{i'j'k'l'}' ~\left(\nabla'{\mathbf u'}\right)_{k'l'} +S_{i'j'l'} {u}_{l'}'  \; , \;\\
 p_{j'}'= {D}_{j'k'l'} \left(\nabla'{\mathbf u'}\right)_{k'l'}  - \sqrt{-1}~\omega~ {\rho'}_{j'l'}{u'}_{l'} \; ,
\\
\label{navier4}
\end{array}
\end{eqnarray}
with $i',j',k',l'=r',\theta',\phi'  $ and  $\nabla'$  the transformed gradient written in transformed coordinates.
One notes that the transformed stress  ${\boldsymbol{\sigma}'}$ is generally not symmetric and the density $\rho'$ is now a second order tensor.
In order to preserve
the symmetry of the stress tensor, one  has to
assume  \cite{norris2011} that  ${\bf A}$ is a multiple of the Jacobian matrix $\partial{\bf x}'/\partial{\bf x}$ of the transformation, i.e  ${ A_{ij}}=\xi \left(\partial{\bf x}'/\partial{\bf x}\right)_{ij}$ ,
where $\xi$ is a non-zero scalar, in which case one obtains a  Willis-type equation \cite{willis1981,milton2006}.
However, another  special case of ${\bf A}$  for which  the elasticity tensor
${\C}'$ does not have the minor symmetries (Cosserat Material \cite{cosserat}), but the stretched density
$\rho'$ is now
a scalar field (unlike in the work by \cite{milton2006}),  is when  ${\bf A}$ is the identity matrix ${\bf I},$ which leads to  \cite{brunapl,diatta-guenneauAPL2014}~{\bf u}'={\bf u}  and 
\begin{eqnarray}
\begin{array}{ll}
\sigma_{i'j'}'=C_{i'j'k'l'}' ~\left(\nabla'{\mathbf u'}\right)_{k'l'}  \; , \;\\
 p_{j'}'=   - \sqrt{-1}~\omega~ {\rho'}{u'}_{j'} \; ,
\\
\label{navier5}
\end{array}
\end{eqnarray}
where the body force is assumed to be zero.

 This equation is derived from (\ref{navier4}) by
noting that ${\bf S}={\bf D}={\bf 0}$  when
${\bf A}={\bf I}$, see \cite{norris2011}. 
In the sequel, we work in the framework of  (\ref{navier3}) coupled with (\ref{navier5}).

\section{Kohn's transform for non singular elastic cloaks}
\begin{figure}[!htb]
\resizebox{120mm}{!}{\includegraphics{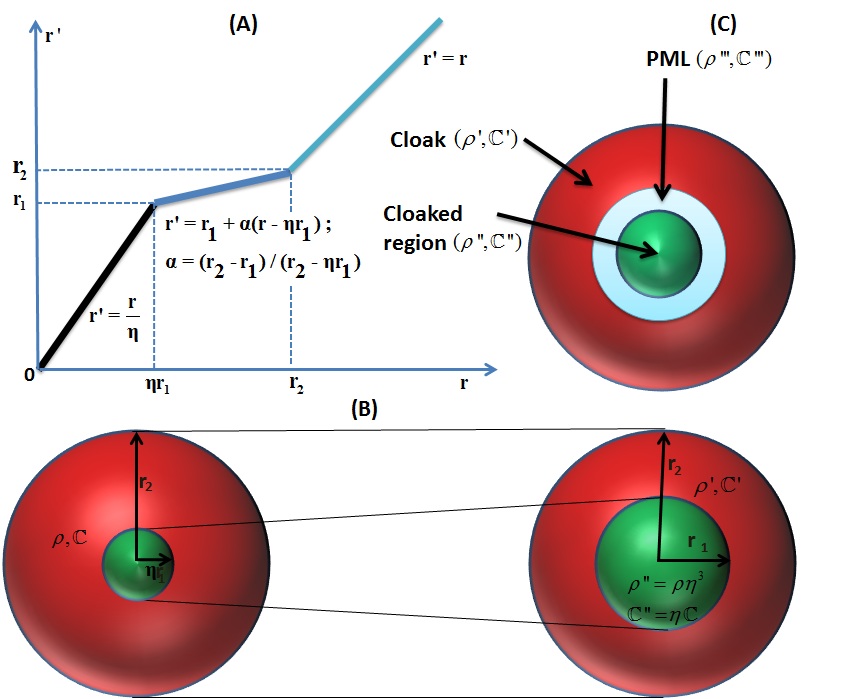}}
\caption{Illustration of geometric transform (\ref{PTransform1}) with (A) its graph
as a broken line and (B) the
subsequent stretch of elastic space from ${\bf x}=(r,\theta,\phi)$ (left)
to ${\bf x}'=(r',\theta',\phi')$ (right), where $r_1$ and $r_2$ are the inner and outer
radii of the spherical cloak, respectively. In terms of material parameters 
a small sphere of radius $\eta r_1$ coated with a hollow sphere of outer radius $r_2$ (left) both of which with an isotropic homogeneous symmetric elastic constitutive
tensor $\C$ and a homogeneous scalar density $\rho$, are mapped onto a larger sphere of radius $r_1$ with an isotropic homogeneous symmetric elastic constitutive
tensor $\C''=\eta\C$ and a homogeneous scalar density $\rho''=\rho\eta^3$, which is coated by a hollow sphere of radius $r_2$ with a heterogeneous asymmetric elastic constitutive tensor $\C'$, see (\ref{sc}) and a heterogeneous scalar density $\rho'$, see (\ref{srho}). In (C), a schematic picture of the construction of a mixed cloak is shown: it is based on the cloak in (B) however with an inner boundary which has a PML attached to it in order to smear down residual elastodynamic energy within the invisibility region.  Note that the inner sphere (cloaked region) consists of an isotropic homogeneous elastic medium with a symmetric elastic constitutive tensor $\C''$, whereas both PML and cloak have asymmetric heterogeneous elasticity tensors $\C'''$ and $\C'$ and heterogeneous scalar densities $\rho'''$ and $\rho'$.
}
\label{3d_cloak_sphere_elastic1}
\end{figure}

Let us now consider the geometric transform,
\begin{eqnarray}
r'= 
\left\{
\begin{array}{lr}
\frac{r}{\eta},
 \label{PTransform1}        \text{if  }  0\leq r\leq\eta r_1
\\
r_1+\frac{r_2-r_1}{r_2-\eta r_1}\left(r-\eta r_1\right) , \label{PTransform2}   
    \text{ if  } \eta r_1\leq r\leq r_2
 \\
r   \text{,  if  } r \ge r_2
\end{array}
\right.
\end{eqnarray}
 which was first
introduced in the context of cloaking in the conductivity equation \cite{kohn2},  where
$
 \theta'=\theta
\; , \; \phi'=\phi
$. 
This geometric transform maps, in a  one-to-one smooth way, a sphere  $0\leq r\leq\eta r_1$ of radius $\eta r_1$  onto the sphere  $0\leq r\leq r_1$  and  the shell  $\eta r_1\leq r'\leq r_2$,  onto the shell
$r_1\leq r'\leq r_2$, as illustrated in  Figure \ref{3d_cloak_sphere_elastic1}-(A) and  Figure \ref{3d_cloak_sphere_elastic1}-(B). In the cylindrical case, design of transformation-based Cosserat elastic cloaks has been
first discussed in \cite{brunapl} , where it only involved a tensor
$\C'$ with $8$ non-vanishing coefficients. In the present spherical
case, we need to consider a tensor $\C'$ with $21$ 
non-vanishing coefficients \cite{diatta-guenneauAPL2014}. Moreover, the displacement field
has three components in our case.
By application of transformation (\ref{PTransform1}) in the region
$ 0\leq r\leq r_2,$ the Navier equations (\ref{navier1}) together with relations (\ref{navier2}) are mapped onto the
equations (\ref{navier3}) coupled with (\ref{navier5}), with
\begin{eqnarray}
\rho'= ab^2(r') \,\rho \; , \label{srho}
\end{eqnarray}
where $a=\frac{r_2-\eta r_1}{r_2-r_1}$ and $b(r')=\frac{(r'-r_1)}{r'}a +\frac{\eta r_1}{r'}$,
and the elasticity tensor ${\bf \C}'$ has 21 non-zero spherical components (it is illuminating to compare these coefficients with those of a fully symmetric elasticity
tensor for an isotropic homogeneous
medium that we recall in Appendix A, see Eq. \ref{sc3}), namely
\begin{eqnarray}
C_{r'r'r'r'}'= (\lambda+2\mu)\frac{b^2(r')}{a},  
C_{\theta\theta\theta\theta}'=C_{\phi \phi\phi\phi}'=  (\lambda+2\mu)a,
\nonumber \\
C_{r'r'\theta\theta}'=  C_{\theta\theta r'r'}'=C_{r'r'\phi\phi}'= C_{\phi \phi r'r'}'=\lambda b(r'), \nonumber \\
C_{\theta\theta\phi\phi}'=  C_{\phi\phi\theta\theta}'=\lambda a, ~
C_{r'\theta r'\theta}'=  C_{r'\phi r'\phi}'=\frac{b^2(r')}{a}\mu, \nonumber \\
C_{\theta r'\theta r'}'=  C_{\theta\phi \theta\phi}'=C_{\phi r' \phi r' }'=  C_{\phi\theta\phi \theta}'=a\mu, \nonumber
\\
C_{r'\theta\theta r'}'=  C_{\theta r'r'\theta}=  C_{r'\phi\phi r'}'= C_{\phi r'r' \phi }'=b(r')\mu, \nonumber 
\\
C_{\theta\phi \phi\theta}'=C_{\phi\theta\theta \phi}'=a\mu \;  . \;
\label{sc}
\end{eqnarray}
Inside the resulting ball of radius $r_1$, the density  and elasticity tensor, here denoted by $\rho''$ and $\C''$ respectively, are obtained from formulas (\ref{srho})-(\ref{sc}) by setting  $ a=\eta=b.$ Namely, 
\begin{equation}
\rho''= \eta^3 \,\rho \; , \label{srho1}
\end{equation}
and   ${\bf \C}'' =\eta\C,$ that is ${\bf \C}''$   is isotropic, homogeneous and fully symmetric, with 21 non-zero spherical components: 
\begin{eqnarray}\label{eq:tensorincentre}
{C}_{r'r'r'r'}''=
{C}_{\theta\theta\theta\theta}''={C}_{\phi \phi\phi\phi}''=  (\lambda+2\mu) \eta,
\nonumber \\
{C}_{r'r'\theta\theta}''=  {C}_{\theta\theta r'r'}''={C}_{r'r'\phi\phi}''= {C}_{\phi \phi r'r'}''=
{C}_{\theta\theta\phi\phi}'' \nonumber\\
=  {C}_{\phi\phi\theta\theta}''=\lambda\eta,\nonumber
\\
{C}_{r'\theta r'\theta}''= {C}_{r'\phi r'\phi}''=
{C}_{\theta r'\theta r'}''=  {C}_{\theta\phi \theta\phi}''\nonumber
\\
={C}_{\phi r' \phi r' }''=  {C}_{\phi\theta\phi \theta}''=
{C}_{r'\theta\theta r'}''= {C}_{\theta r'r'\theta}''\nonumber
\\
= {C}_{r'\phi\phi r'}''= {C}_{\phi r'r' \phi }''=
{C}_{\theta\phi \phi\theta}''={C}_{\phi\theta\theta \phi}''=\mu \eta\;  . \;
\label{sc1}
\end{eqnarray}

\section{Physical discussion of the structure of the elasticity tensor}
One notes that when $\eta=0$, (\ref{srho}) and (\ref{sc}) have the same form as in \cite{diatta-guenneauAPL2014} as Kohn's transform \cite{kohn2} coincides with Pendry's transform\cite{pendry} and when $\eta=1$, Kohn's transform reduces to the identity so that (\ref{srho})-(\ref{sc1}) all have the same entries as the elasticity tensor of the surrounding medium.
As discussed above, one should note that the minor symmetries are broken and $b(r_1)=\eta$ which means
$C'_{\theta\theta\theta\theta}/C'_{r'r'r'r'}=C'_{\phi \phi\phi\phi}/C'_{r'r'r'r'}$ is $\frac{1}{\eta^2 }$
at the inner boundary of the cloak (anisotropy of  $\frac{1}{\eta^2 }$), unlike in our
previous work \cite{diatta-guenneauAPL2014}, wherein the anisotropy was infinite
(since $\eta$ was equal to zero). Moreover, the off-diagonal components
are also constant  at the boundary $r'=r_1$, they are nonzero whenever $\eta\neq 0.$
The physical interpretation is that shear and pressure waves propagate much faster (the smaller $\eta$, the faster)  in the azimuthal and elevation directions than in the radial direction on the surface of the inner boundary.
In fact, the waves' acceleration in azimuthal and elevation directions makes possible a vanishing phase shift between an elastic wave propagating in an isotropic homogeneous elastic medium, and another one propagating around the core region
(without this acceleration, it is clear that the longer wave trajectory in the latter case would induce a deformed wavefront in forward scattering, see e.g. the shadow region in the lower panels of Figure \ref{3d_cloak_sphere_elastic_mesh}).
One should also note that there are no infinite entries within the elasticity tensor, unlike for the cylindrical cloak studied in \cite{brunapl}.
Let us further note that comparing (\ref{sc}) and (\ref{sc1}) at $r'=r_1$,  for $i,j=r',\theta,\phi$ and $i\neq j,$ we have  $C_{iijj}'=C_{iijj}''=\lambda\eta$ and  $C_{ijji}'=C_{ijij}'=C_{ijji}''=C_{ijij}''=\mu\eta,$ 
 which means that these entries of the transformed elasticity tensor are continuous across the interface $r'=r_1$. However, $C'_{rrrr}/{C}_{rrrr}''=\eta/a$ and $C_{\theta\theta\theta\theta}'/{C}_{\theta\theta\theta\theta}''=C_{\phi\phi\phi\phi}'/{C}_{\phi\phi\phi\phi}''=a/\eta$ 
 ($\C''$ is the tensor in the inner ball). This is in accordance with the fact that the elasticity tensor $\C'$  is strongly anisotropic in the azimuthal and elevation directions, whereas $\C''$ is isotropic and has entries of order $\eta$.

Let us now note that when $r=r_2$, the geometric transform
(\ref{PTransform1}) leads to $r'=r_2$. In this case, the transformed
density is $\rho'=a\rho$ as $b(r_2)=1$ and the diagonal
components of the transformed
elasticity tensor reduce to
$
C'_{r'r'r'r'}=(\lambda+2\mu)/a \; ,
C'_{\theta\theta\theta\theta}=
C'_{\phi\phi\phi\phi}
=(\lambda+2\mu)a \; ,
$
hence $C'_{r'r'r'r'}=C_{rrrr}/a$,
$C'_{\theta\theta\theta\theta}=a C_{\theta\theta\theta\theta}$
and
$C'_{\phi\phi\phi\phi}=a C_{\phi\phi\phi\phi}$
on the outer boundary of the cloak.
This expresses the fact that a stretch along the radial direction
is compensated by a contraction along the azimuthal
and elevation directions in such a way that elastic media (cloak and surrounding isotropic elastic medium) are
impedance-matched at $r'=r_2$.
The cloak's outer boundary therefore behaves in many ways as an
impedance matched `thin' elastic layer.
However, we note that the components of ${\C}'$ pose no limitations
on the applied frequency $\omega$ from low to high frequency, unlike for the
case of coated cylinders studied back in 1998 in the context of elastic
neutrality by Bigoni et al. \cite{bigoni98}.
The fact that ${\C}'$ does not depend on $\omega,$ i.e. the cloak consists
of a non-dispersive elastic medium, makes it work at all frequencies, but
one should keep in mind that any structured medium designed to approximate
the ideal cloak's parameters (e.g. via homogenization) would necessarily
involve some dispersion, and thus limit the interval of frequencies over
which the cloak can work. Such a feature has been already observed
in \cite{farhat2012,stenger2012}, for cloaking of flexural waves in thin-elastic plates.

\section{Numerical implementation and  illustrations}
Let us now numerically investigate the cloaking efficiency. In order to do this, we implement the $3^4$ spatially varying
entries of the transformed tensor in Cartesian coordinates in the finite element package COMSOL MULTIPHYSICS.
We mesh the computational domain using 1105932 tetrahedral elements, 36492 triangular elements, 1128 edge
elements and 25 vertex elements.
This domain consists of  an isotropic homogeneous elastic medium within a sphere of radius $r_3=10$~m, containing  a small sphere (an isotropic homogeneous medium) of radius $r_1=2$~m surrounded by the cloak (a heterogeneous, anisotropic elastic metamaterial in a spherical shell of inner radius $r_1=2$~m and outer radius $r_2=4$~m ) 
and  a point force oriented along the direction $(1,1,1)$, which is located at $(4 ~ m,-7 ~m,0~m),$ where $(0,0,0)$ is the center of the cloak. 
The sphere of radius $r_3=10$~m is itself surrounded by a spherical shell of inner radius $r_3$ and outer radius $r_4=12$~m, 
which is filled with an anisotropic homogeneous absorptive
 medium of Cosserat type  acting as a (reflectionless) perfectly matched layer (PML).

\subsection{Elastic spherical perfectly matched layers}

In this section,  we supply a fully detailed (i.e. complete) expression for the spherical elastic Perfecttly Matched Layers of Cosserat type. 
 We consider the radial function $s_r(r)=1-i$  and deduce elastic spherical PML from the  geometric transform
\begin{equation}
r''=r_3+\displaystyle\int_{r_3}^{r}s_r(v)dv \; , \; \theta''=\theta
\; , \; \phi''=\phi \; . \label{PML}
\end{equation}
This transform leads, in the same way as (\ref{PTransform2}) did for $\C'$ and $\rho'$, to a homogeneous anisotropic asymmetric elasticity tensor $\C'''$ and a scalar (homogeneous) density $\rho''',$ with 
\begin{eqnarray}
C_{r'r'r'r'}'''=\frac{(1+i)}{2} (\lambda+2\mu),  
C_{\theta\theta\theta\theta}'''=C_{\phi \phi\phi\phi}'''= (1-i) (\lambda+2\mu),
\nonumber \\
C_{r'r'\theta\theta}'''=  C_{\theta\theta r'r'}'''=C_{r'r'\phi\phi}'''= C_{\phi \phi r'r'}'''=\lambda, \nonumber \\
C_{\theta\theta\phi\phi}'''=  C_{\phi\phi\theta\theta}'''= (1-i) \lambda, ~
C_{r'\theta r'\theta}'''=  C_{r'\phi r'\phi}'''=\frac{(1+i)}{2}\mu, \nonumber \\
C_{\theta r'\theta r'}'''=  C_{\theta\phi \theta\phi}'''=C_{\phi r' \phi r' }'''=  C_{\phi\theta\phi \theta}'''= (1-i)\mu, \nonumber
\\
C_{r'\theta\theta r'}'''=  C_{\theta r'r'\theta}'''=  C_{r'\phi\phi r'}'''= C_{\phi r'r' \phi }'''=\mu, ~~
C_{\theta\phi \phi\theta}'''=C_{\phi\theta\theta \phi}'''=(1-i)\mu \;  . \;
\label{scpml}
\end{eqnarray}
where, again,  $\lambda$ and  $\mu$  are the Lam\'e coefficients of the ambient homogeneous and isotropic space, and $\rho'''=(1-i) \rho$.
Since the elastic waves are damped inside the PML and reach the outer boundary of the shell with a vanishing amplitude, we set either clamped or traction free boundary conditions at $r''=r_4$ (we have checked this does not affect the numerical result).

\subsection{Finite element results}

\begin{figure}[!htb]
\resizebox{84mm}{!}{\includegraphics{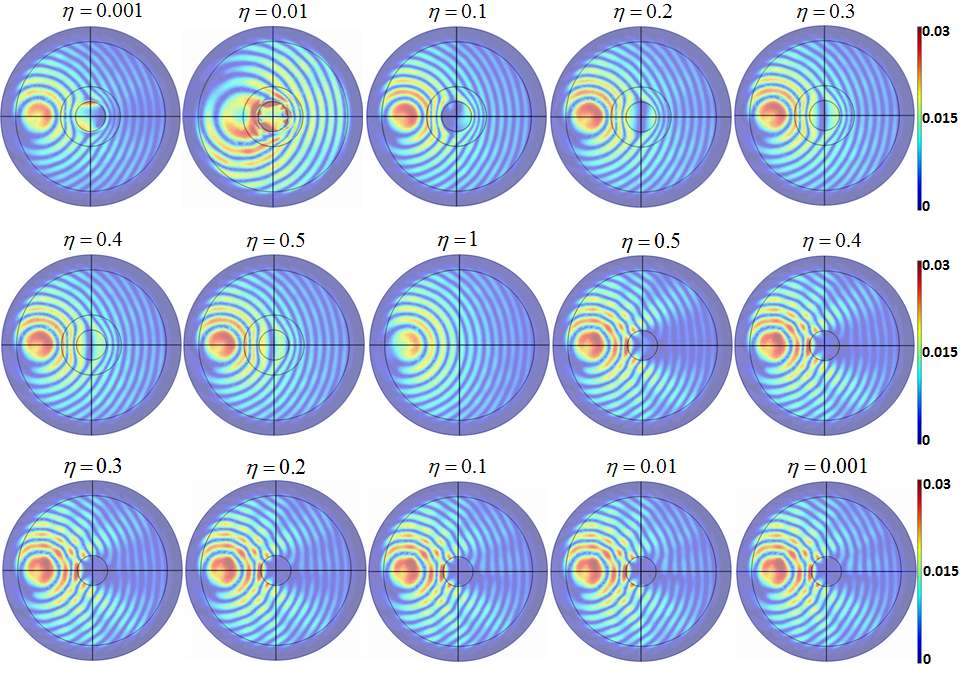}} 
\vspace{-3mm}
\caption{
Spherical cloak (upper row and the first two images to the left of the middle row)  with elastic parameters as in (\ref{sc}) and (\ref{srho}) of outer radius $4$m dressing a soft sphere of radius $2$m with Lam\'e parameters $\mu=\eta$, $\lambda=2.3\eta$ and $\rho=\eta^3$, with $\eta=0.001, ~ 0.01, ~0.1,  ~0.2, ~ 0.3, ~0.4, ~0.5$ and  the same soft sphere alone (third row and the last two images of the middle row) within a homogeneous isotropic medium with Lam\'e parameters $\mu=1$, $\lambda=2.3$ and $\rho=1$.  Numerical illustration of the real part of complex valued magnitude $\sqrt{u_1^2+u_2^2+u_3^2}$ of the elastic displacement, for a time-harmonic point source oriented along $(1,1,1)$ at the frequency $\omega=3$ rad.s$^{-1}$ located at a distance $r=8.062$m from the center of  the soft sphere. Note that the isotropic homogeneous elastic material  in the central panel ($\eta=1$) serves as a benchmark.}
\label{3d_cloak_sphere_elastic_mesh}
\end{figure}

One should keep in mind that any numerical implementation in a finite element package requires a Cartesian coordinate system. In our case, we used COMSOL where the transformed elastic tensors $\C'$ for the cloak and $\C'''$ for the PML had up to $81$ non-vanishing spatially varying entries. A good way to detect any flaw in the numerical implementation is to compare the solution to the
problem for a time harmonic point source in an isotropic homogeneous medium (supplied with spherical elastic PML),
see the center panel for $\eta=1$ in Figure \ref{3d_cloak_sphere_elastic_mesh}, with
the solution to the same problem when we have cloaks surrounding certain soft spheres, as shown in other panels
in  Figure \ref{3d_cloak_sphere_elastic_mesh}, where we vary the $\eta$ parameter in the range $0.001$ (extremely soft sphere and cloak with nearly singular strongly anisotropic elasticity tensor with marked minor symmetry breaking) to $0.5$ (mildly soft sphere with weakly anisotropic and asymmetric elasticity tensor). The magnitude of the elastic displacement outside the cloaks is virtually
indistinguishable from that of the center panel (case $\eta=1$).
\begin{figure*}[!htb]
\resizebox{100mm}{!}{\includegraphics{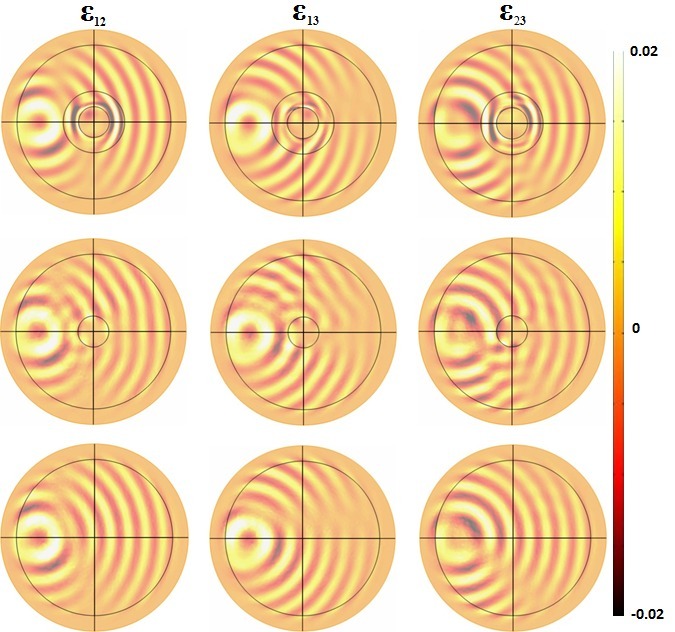}}
\vspace{-3mm}
\caption{
Deformation of a homogeneous isotropic elastic medium $\mu=1$, $\lambda=2.3$ and $\rho=1$ (lower panel); Deformation of the same medium now containing a soft homogeneous isotropic spherical obstacle of radius $2$ m with Lam\'e parameters $\mu=\eta$, $\lambda=2.3\eta$ and $\rho=\eta^3$ for $\eta=0.1$ (middle panel); Deformation of the same obstacle but now surrounded by a heterogeneous elastic cloak with an elasticity tensor without the minor symmetries (upper panel) of inner radius $r_1=2$~m and outer radius $r_2=4$~m, subjected to a concentrated load of polarization $(1, 1, 1)$ with frequency  $\omega=3$ rad.s$^{-1}$ located in a homogeneous isotropic medium (cf. upper panel) at a distance $8.062$~m from the origin. Real part of the deformation components
 $\varepsilon_{12}=\frac{1}{2}(\frac{\partial u_1}{\partial x_2}+\frac{\partial u_2}{\partial x_1})$
(left column), $\varepsilon_{13}=\frac{1}{2}(\frac{\partial u_1}{\partial x_3}+\frac{\partial u_3}{\partial x_1})$   (middle column) and
$\varepsilon_{23}= \frac{1}{2}(\frac{\partial u_2}{\partial x_3}+\frac{\partial u_3}{\partial x_2})$ (right column) of the strain tensor $\varepsilon$.
}
\label{3d_cloak_sphere_elastic3a}
\end{figure*}

\begin{figure}[!htb]
\resizebox{85mm}{!}{\includegraphics{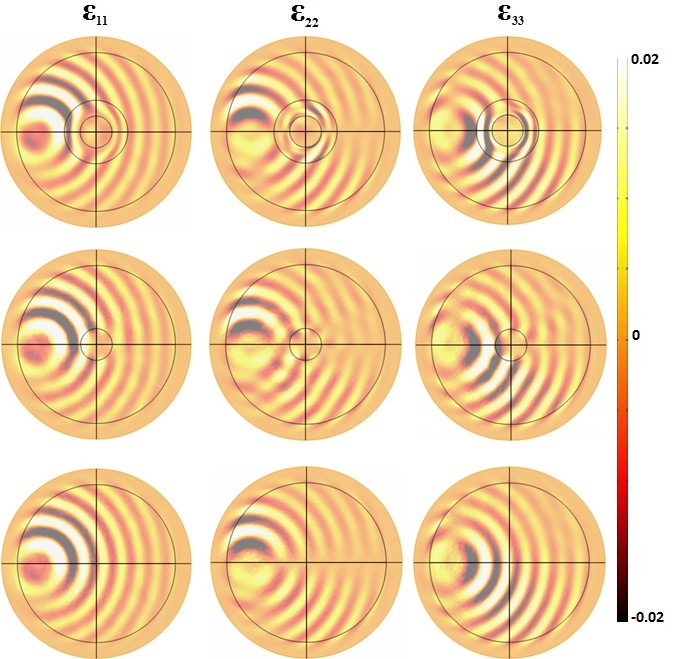}}
\vspace{-3mm}
\caption{
Deformation of an isotropic elastic medium (lower row), a soft spherical obstacle (middle row) and the same obstacle surrounded by an elastic cloak (upper row) for the same geometrical and elastic parameters and the same source (and frequency) as in Fig. \ref{3d_cloak_sphere_elastic3a}. Real part of the deformation components $\varepsilon_{11}=\frac{\partial u_1}{\partial x_1}$
(left column), $\varepsilon_{22}=\frac{\partial u_2}{\partial x_2}$  (middle column) and
$\varepsilon_{33}=\frac{\partial u_3}{\partial x_3}$ (right column) of the strain tensor $\varepsilon$.
}
\label{3d_cloak_sphere_elastic3b}
\end{figure}
For $\omega=3$ rad.s$^{-1}$ and $\eta=0.1,$ one can see in Figure \ref{3d_cloak_sphere_elastic3a} and Figure \ref{3d_cloak_sphere_elastic3b} that the deformation of the
elastic medium outside the cloak is nearly identical to that of
the isotropic homogeneous elastic medium (we use the normalized density $\rho=1$
and Lam\'e parameters $\mu=1$ and $\lambda=2.3$, for respectively the shear modulus and the compressibility).
By comparison, the deformation of the elastic medium is clearly visible for the soft sphere (of normalized density $\rho=\eta^3$
and Lam\'e parameters $\mu=\eta$ and $\lambda=2.3\eta$) without the cloak.

 The small discrepancy between the upper and lower panels in Figure  \ref{3d_cloak_sphere_elastic3a}, third column, is attributed to the artificial anisotropy induced by the mesh (it is not radially symmetric, see \cite{diatta-guenneauAPL2014}) and the small absorption due to PMLs.

\begin{figure*}[!htb]
\resizebox{120mm*}{!}{\includegraphics{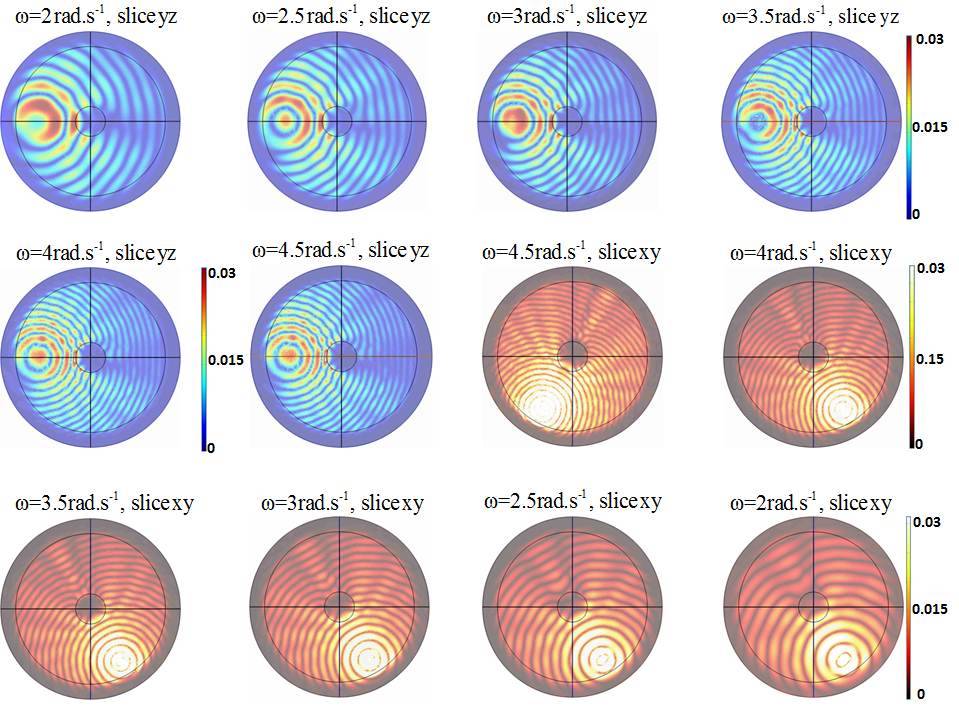}}
\vspace{-3mm}
\caption[!htb]{Real part of complex valued magnitude $\sqrt{u_1^2+u_2^2+u_3^2}$ of the elastic displacement for a time-harmonic point source oriented along the direction (1,1,1), in the frequency range $\omega=2$ to $4.5$ rad.s$^{-1}$, located at a distance $r=8.062$m from the center of a spherical soft sphere (without a surrounding cloak), with parameters as in Fig. \ref{3d_cloak_sphere_elastic3a}. 
}
\label{3d_cloak_sphere_elastic3}
\end{figure*}

\begin{figure}[!htb]
\resizebox{120mm}{!}{\includegraphics{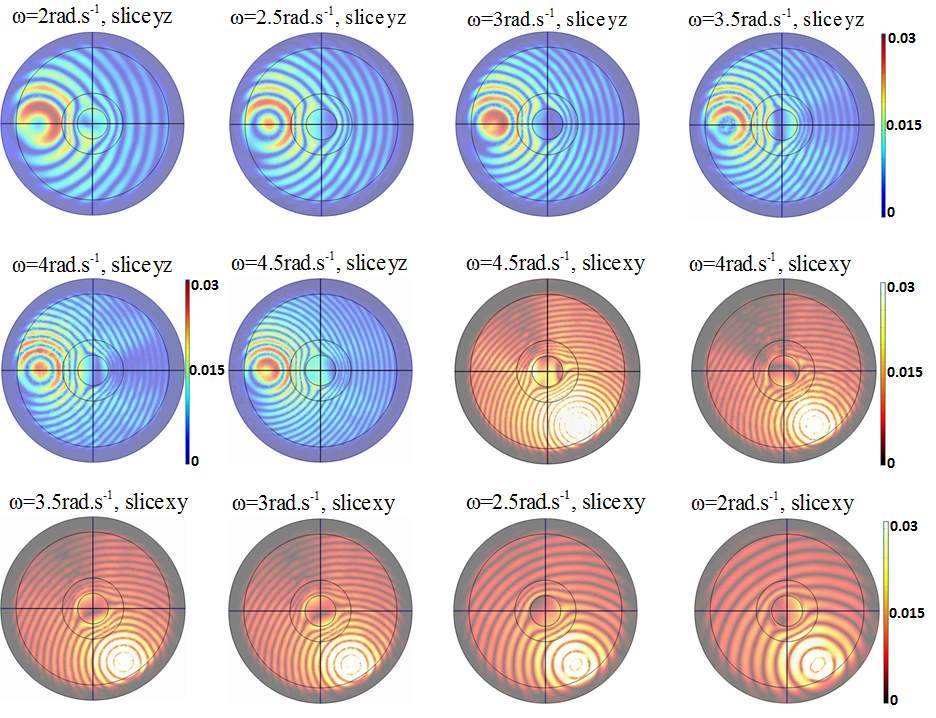}}
\vspace{-3mm}
\caption{Real part of complex valued magnitude $\sqrt{u_1^2+u_2^2+u_3^2}$  of the elastic displacement for a time-harmonic point source in the frequency range $\omega=2$ to $4.5$ rad.s$^{-1}$ located at a distance $r=8.062$m from the center of a spherical soft sphere and oriented along $(1,1,1),$ with a surrounding cloak with parameters as in Fig. \ref{3d_cloak_sphere_elastic3a}.  
}
\label{3d_cloak_sphere_elastic2}
\end{figure}

 Plots of elastic deformation can be somewhat misleading, and we therefore add plots of the magnitude of the elastic displacement field, for $\eta=0.1,$ in Figure  \ref{3d_cloak_sphere_elastic3} (scattering by a soft sphere)
 and Figure \ref{3d_cloak_sphere_elastic2} (scattering by a soft sphere surrounded by the cloak) both
for angular frequencies varying in the range $2$ to $4.5$ rad.s$^{-1}$, where it should be noticed that the shaded region behind the soft sphere in Figure  \ref{3d_cloak_sphere_elastic3} (drop of wave amplitude and phase shift), is almost completely removed in  Figure \ref{3d_cloak_sphere_elastic2} thanks to the cloak. Upon inspection of
these two figures, one can clearly see that the elastic field scattered by a soft sphere clothed with the cloak is virtually indistinguishable from bare isotropic elastic space.

  However, these
results may not be representative of the cloak's behavior at other frequencies. It is also
possible that other parameters $\eta$ lead to cloaks offering a better wave protection. Figure \ref{3d_cloak_sphere_elastic_mesh} shows the magnitude of the displacement field for $\omega=3$ rad.s$^{-1}$
and for 7 values of $\eta$ ranging from $0.001$ to $0.5.$

\begin{figure}[!htb]
\resizebox{120mm}{!}{\includegraphics{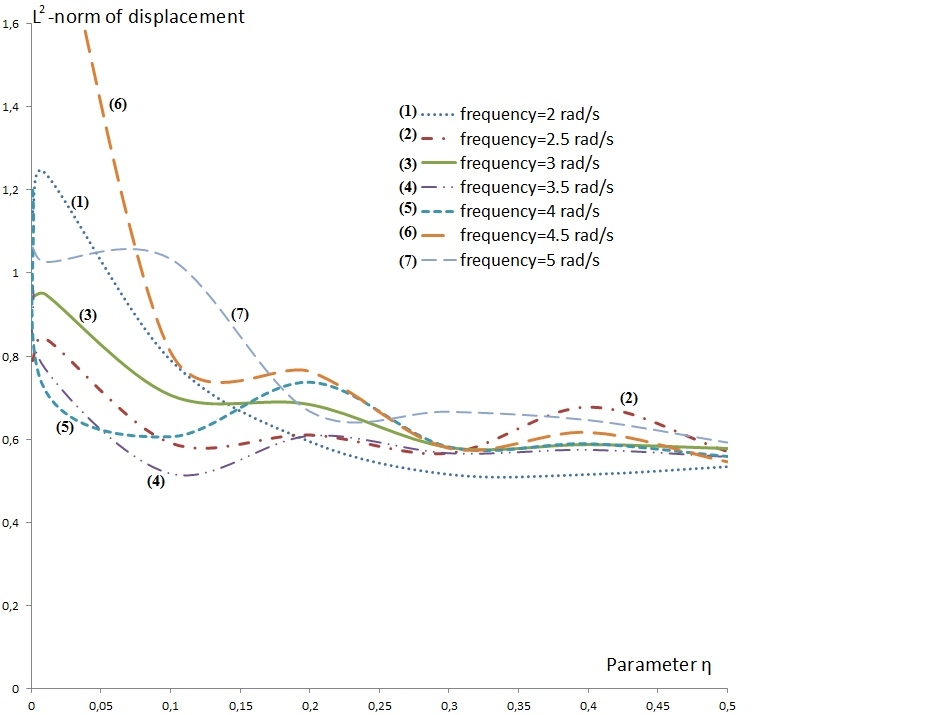}}
\vspace{-3mm}
\caption{ $L^2$-norm $\Vert {\bf u} \Vert_{L^2}=\sqrt{\int(u_1\bar{u_1}+u_2\bar{u_2}+u_3\bar{u_3})}$ of the displacement field inside the small isotropic homogeneous soft balls of radius $r_1=2~m$ surrounded by  cloak of outer radius  $r_2=4~ m$, against the parameter $\eta$ varying from 0.001 to 0.5,  for the same point force as in Fig. \ref{3d_cloak_sphere_elastic_mesh}.
}
\label{3d_cloak_sphere_elastic4}
\end{figure}

It seems plausible that the elastic wave displacement
vanishes inside the cloaked region, but this is in contradistinction with plots of elastic field magnitude
in Figures  \ref{3d_cloak_sphere_elastic2}  and \ref{3d_cloak_sphere_elastic_mesh}. It is also important to have some quantitative criterion regarding the possible wave protection inside the
cloaked region (the soft sphere within the cloak).
In order to check
more thoroughly what is the level of wave protection depending upon the wave frequency and
the value of $\eta$ (in a two-parameter space), we plot in Figure \ref{3d_cloak_sphere_elastic4}
the $L^2$-norm of the elastic displacement field within the sphere
for given frequencies (see curves 1 to 7) against the parameter $\eta$. We conclude that the most
favorable pair of parameters lies within an interval centered at $\omega=3$ rad.s$^{-1}$ and $\eta=0.1$, which 	are indeed the
values used in Figures \ref{3d_cloak_sphere_elastic3a} to  \ref{3d_cloak_sphere_elastic_mesh}. 
A  further  quantitative investigation of the favorable  case  $\eta=0.1$,  still exhibits some large peaks of values of the $L^2$-norm of the displacement field within the cloaked soft sphere, yet due to some resonances, as shown  in  Figure \ref{3d_cloak_sphere_elastic5}. In the same vein, Figures  \ref{3d_cloak_sphere_elastic2}   and \ref{3d_cloak_sphere_elastic_mesh}, show  qualitative studies  for the case $\eta=0.1$ ($\omega$ varying) and $\omega=3$ rad.s$^{-1}$ ($\eta$ varying), respectively.
 We can conclusively say that our cloak works well in terms of invisibility but does not offer a wave protection at all frequencies. 
We shall indeed see in the next section \ref{sect:resonances-approximation}, that the cloak actually makes a concentrator for elastic fields at specific frequencies. In section \ref{sect:mixedcloaks}  we shall propose a new generation of cloaks, termed here mixed cloaks, that  offer a more accomplished wave protection. 

\section{Estimate of local resonance inside the cloak}\label{sect:resonances-approximation}
\begin{figure}[!htb]
\resizebox{85mm}{!}{\includegraphics{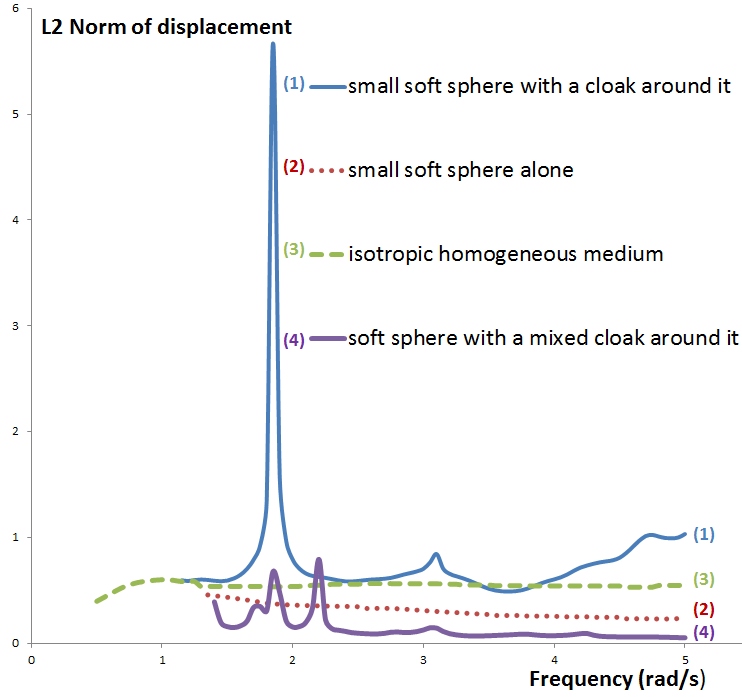}}
\vspace{-3mm}
\caption{
$L^2$-norm $\Vert {\bf u} \Vert_{L^2}=\sqrt{\int_{B({\bf 0}, r_1)} (u_1\bar{u_1}+u_2\bar{u_2}+u_3\bar{u_3})}$ 
of the displacement field 
inside the small soft ball of radius $r_1=2~m$,  for the fixed value $\eta=0.1,$
when (1) surrounded by the cloak of outer radius $r_2=4~ m$,   (2) alone,  (3) when  its elastic properties are replaced by those of the ambient isotropic homogeneous elastic medium, and  (4) when it is now surrounded by a mixed cloak (made of 2 heterogeneous anisotropic layers: the same cloak as in (1) completed by a spherical  region of inner radius $r_1=1.5$~m and outer radius $r_2=2$~m containing  PML), the $L^2$ being computed over the sphere of radius $2$ m (PML+ soft obstacle). The frequency varies by 0.05 rad.s$^{-1}.$ The point force is as in  Fig. \ref{3d_cloak_sphere_elastic_mesh}.
The cloaked sphere being soft allows for a better highlight of the existence of resonance peaks in the case of an ordinary cloak (1). This clearly demonstrates the lack of wave protection within the cloaked soft sphere, for some resonances. However, when the same soft sphere is surrounded by a mixed cloak (4), the large resonance (1) is reduced by a factor greater than 6. Note also that  a good wave protection is achieved for all other frequencies.
}
\label{3d_cloak_sphere_elastic5} 
\end{figure}

We would like to gain some physical insight in trapped eigenstates within the core of the cloaks, which play an antagonistic role in seismic wave protection, as it transpires in Figure \ref{3d_cloak_sphere_elastic2}. To do this, the governing equations associated with the invisibility cloak should be simplified, and a way to achieve this is by assuming
that the Lam\'e parameter $\lambda=0$ (vanishing compressibility modulus) and ${\bf u}={\bf u}(r)$ (this seems legitimate in view of the fact that the problem is radially symmetric), then the three components of the displacement field are all solutions of the scalar Helmholtz equation in spherical coordinates:
\begin{equation}
\begin{array}{lll}
&\displaystyle{\frac{\partial}{\partial r}\left(r^2\mu_r\frac{\partial}{\partial r}u_i\right)+\frac{\mu_\theta}{\sin\theta}\frac{\partial}{\partial \theta}\left(\sin\theta\frac{\partial}{\partial \theta}u_i\right)} \\
&+\displaystyle{\frac{\mu_\phi}{\sin^2\theta}\frac{\partial^2}{\partial^2 \phi}u_i+\omega^2\rho u_i=0}
\end{array}
\label{bleu}
\end{equation}
where $\mu_r(r)=\mu_0{(r-r_1)}^2/r^2$, $\mu_\theta(r)=\mu_\phi(r)=\mu_0 r_2/(r_2-r_1)$ and $\rho(r)=\rho_0({(r-r_1)}^2/r^2) (r_2^3/{(r_2-r_1)}^3)$.
The general solution of (\ref{bleu}) is expressed in terms of spherical Bessel functions of first and second kinds:
\begin{equation}
u_i(r)=AJ_0\left(\frac{(r-r_1)\omega\rho_0 r_2^3}{\mu_0{(r_2-r_1)}^3}\right)+BY_0\left(\frac{(r-r_1)\omega\rho_0 r_2^3}{\mu_0{(r_2-r_1)}^3}\right) \; .
\label{solutionparticuliere}
\end{equation}
It remains to supply the simplified governing equation (\ref{bleu}) with two boundary conditions to find a particular solution (i.e. estimate the two constants A and B in (\ref{solutionparticuliere})). At this stage, we use our physical intuition and make the
hypothesis that some of the trapped eigenstates in Figures \ref{3d_cloak_sphere_elastic2} and \ref{3d_cloak_sphere_elastic_mesh} can be approximated by a simple spring-mass model: Upon resonance, the shell of the cloak behaves like an effective spring, connected to a stress-free cavity (a mass) at the inner boundary $r_1$ and to a clamped wall at the outer boundary $r_2$. This effective model is inspired by almost trapped eigenstates unveiled in \cite{greenleaf1,greenleaf2} for Neumann-type cavities in the context of quantum cloaking (Schr\" odinger operator). Then, assuming that $du_i(r_1)/dr=0$ (stress-free boundary condition at the inner boundary of the cloak i.e. freely moving body inside the core) and $u_i(r_2)=0$ (clamped boundary condition at the outer boundary of the cloak i.e. vanishing displacement field outside the cloak) for all three components, we find that the eigenfrequency should
satisfy the following equation
\begin{equation}\label{resonances}
Y_0(\omega r_2 \sqrt{\mu_0/\rho_0})/J_0(\omega r_2 \sqrt{\mu_0/\rho_0})=0 \; ,
\end{equation}
which gives the frequency estimate $\omega\sim 1.52$ rad.s$^{-1}$ for $\mu_0=\rho_0=1$ since the first zero of $Y_0(x)$ is $3.8317$. This frequency estimate is in reasonable agreement with
the finite elements' solution $\omega=1.85$ rad.s$^{-1}$ for the first resonance, which corresponds to the large elastic field enhancement shown
in Fig. \ref{3d_cloak_sphere_elastic5}. The frequency estimate for the second resonance is  $\omega\sim 2.79$ rad.s$^{-1}$ since the second zero of $Y_0(x)$ is $7.0156$
and this is also in reasonable agreement with the finite element's solution $3.1$ rad.s$^{-1}$, which corresponds to the second peak in Fig. \ref{3d_cloak_sphere_elastic5}.
The discrepancy between asymptotics and numerics could also be attributed to some unavoidable shift in resonances induced by the absorption within the PML (wherein
the elastic parameters have a small imaginary part in order to damp the outgoing waves). Indeed, one should note that in Fig. \ref{3d_cloak_sphere_elastic4}
and \ref{3d_cloak_sphere_elastic5} that we compute a real valued $L^2$-norm of the complex valued elastic displacement ${\bf u}$ by multiplying its components $u_i$ by
their complex conjugates $\bar{u_i}$. However, if we now choose to compute the complex valued quantity ${\cal N}({\bf u})=\int_\Omega\sqrt{u_1^2+u_2^2+u_3^2}$ and then take its real part
 $\Re({\cal N}({\bf u}))$,
then it transpires from Fig. \ref{3d_cloak_sphere_elastic4} and Fig. \ref{3d_cloak_sphere_elastic4appendix} that although the overall patterns of curves are preserved, there is a noticeable shift induced in the frequency
corresponding to the global minimum of elastic displacement norm within the soft sphere for $\eta$ ranging from $0.001$ to $0.5$. Indeed, this
global minimum occurs at $\eta=0.1$ for a frequency around $\omega=3$ rad.s$ ^{-1}$ in Fig. \ref{3d_cloak_sphere_elastic4appendix}, instead of
$\eta=0.1$ for a frequency around $\omega=3.5$ rad.s$ ^{-1}$ in Fig. \ref{3d_cloak_sphere_elastic4} for the real valued $L^2$-norm.
We note again that these local resonances (that were not foreseen in \cite{diatta-guenneauAPL2014} since we considered cloaked voids) are reminiscent of those unveiled in \cite{greenleaf1,greenleaf2} in the context of almost trapped eigenstates in quantum cloaking and sensors.
This suggests some sensing potential in elastic waves.

\section{The mixed cloak}\label{sect:mixedcloaks}
In the previous sections, we have singled out the existence of an enhanced field inside the cloaked area, brought in by some resonances. 
One of the purposes of the use of soft spheres in the present work, is their ability to allow for a better highlight of the existence of resonance peaks and hence  the possible lack of wave protection within the cloaked regions.
Through an investigation of trapped modes in elasticity, some reasonably good asymptotic approximation of such (countable set of) resonances is supplied in Section \ref{sect:resonances-approximation}, as a solution of some transcendental equation. 
In this section, we propose 
a new generation of cloaks, the mixed cloaks, in order to achieve cloaking along with protection for elastic (volume) waves. Such a protection might prove very useful in civil engineering.  Indeed, such mixed cloaks overcome the pitfall of inner resonances (trapped modes) of elastodynamic cloaks that might coincide with eigenfrequencies of the buildings to be protected.

\begin{figure}[!htb]
\resizebox{85mm}{!}{\includegraphics{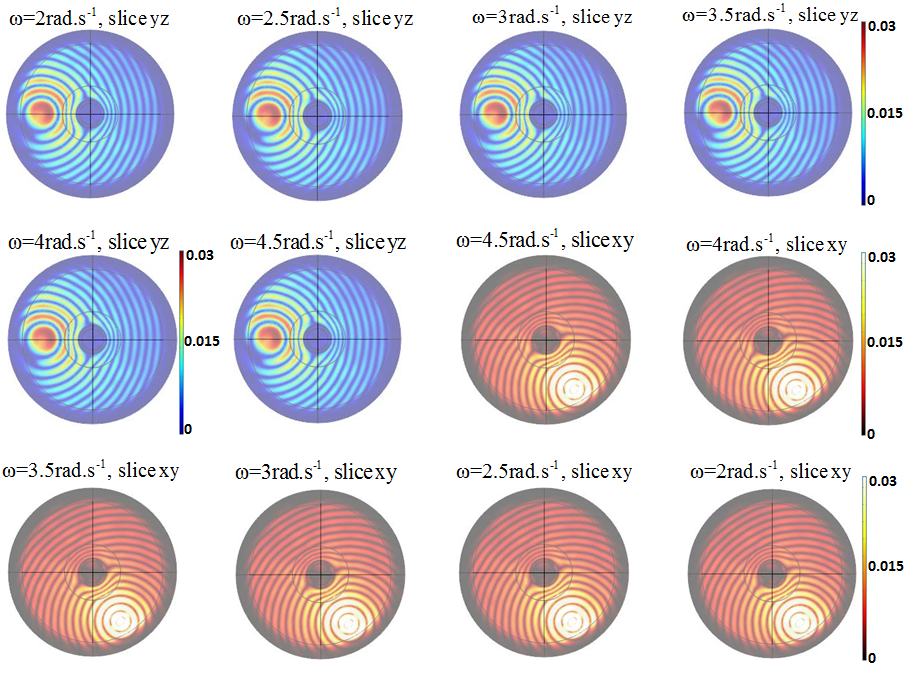}}
\vspace{-3mm}
\caption{Mixed cloak: Real part of complex valued magnitude $\sqrt{u_1^2+u_2^2+u_3^2}$  of the elastic displacement for a time-harmonic point source in the frequency range $\omega=2$ to $4.5$ rad.s$^{-1}$ located at a distance $r=8.062$m from the center of a soft sphere and oriented along $(1,1,1),$ with a surrounding mixed cloak, consisting of an ordinary spherical Cosserat cloak  of inner radius $r_1=2$~m and outer radius $r_2=4$~m and a PML region of inner radius $r_1=1.5$~m and outer radius $r_2=2$~m. The parameters are as in Fig. \ref{3d_cloak_sphere_elastic3a}.   One clearly realizes the absence of field inside the cloaked area, entailing a complete protection for the soft sphere. Compare Fig. \ref{3d_cloak_sphere_elastic2}.
}
\label{3d_cloak_sphere_elastic2_pml}
\end{figure}
\begin{figure*}[!htb]
\resizebox{100mm}{!}{\includegraphics{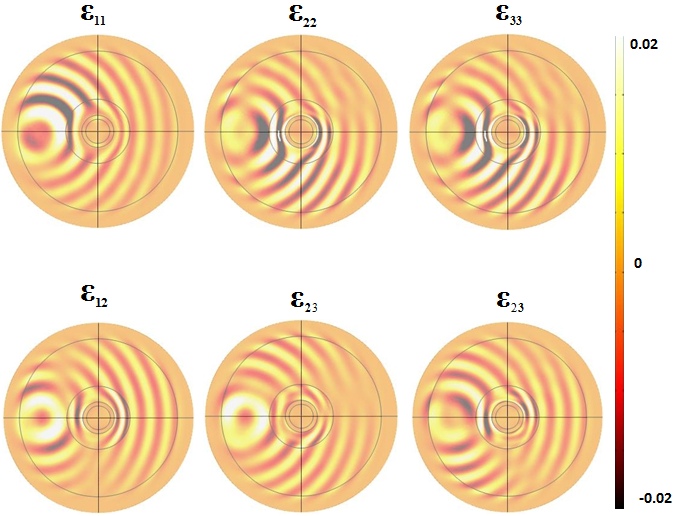}}
\vspace{-3mm}
\caption{Deformation of a homogeneous isotropic elastic medium $\mu=1$, $\lambda=2.3$ and $\rho=1$, containing a soft homogeneous isotropic spherical obstacle of radius $1.5$ m with Lam\'e parameters $\mu=\eta$, $\lambda=2.3\eta$ and $\rho=\eta^3$ for $\eta=0.1$, surrounded by a mixed cloak with an elasticity tensor without the minor symmetries. The mixed cloak is made of an ordinary spherical Cosserat cloak  of inner radius $r_1=2$~m and outer radius $r_2=4$~m, completed by a PML region of inner radius $r_1=1.5$~m and outer radius $r_2=2$~m.  We apply a concentrated load of polarization $(1, 1, 1)$ with frequency  $\omega=3$ rad.s$^{-1}$ located in a homogeneous isotropic medium  at a distance $8.062$~m from the origin. Real part of the deformation components $\varepsilon_{ij}=\frac{1}{2}(\frac{\partial u_i}{\partial x_j}+\frac{\partial u_j}{\partial x_i}),$ with $i,j=1,2,3,$ of the strain tensor $\varepsilon$. Compare  with Figure \ref{3d_cloak_sphere_elastic3a} and Figure \ref{3d_cloak_sphere_elastic3b}.
}
\label{fig:strain_mixed_cloak}
\end{figure*}
\begin{figure*}[!htb]
\resizebox{100mm}{!}{\includegraphics{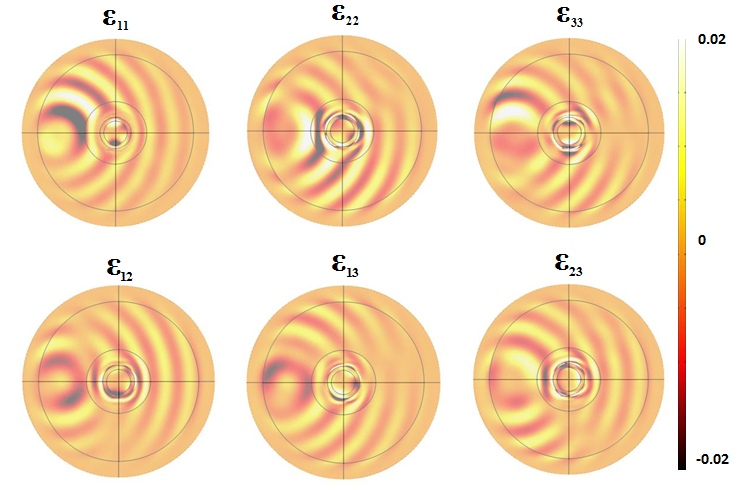}}
\vspace{-3mm}
\caption{Deformation of a homogeneous isotropic elastic medium $\mu=1$, $\lambda=2.3$ and $\rho=1$, containing a soft homogeneous isotropic spherical obstacle of radius $1.5$ m with Lam\'e parameters $\mu=\eta$, $\lambda=2.3\eta$ and $\rho=\eta^3$ for $\eta=0.1$, surrounded by a mixed cloak with an elasticity tensor without the minor symmetries. The mixed cloak is made an ordinary spherical Cosserat cloak  of inner radius $r_1=2$~m and outer radius $r_2=4$~m, completed a PML region of inner radius $r_1=1.5$~m and outer radius $r_2=2$~m.  We apply a concentrated load  (with frequency tuned to the peak $\omega=2.20$ rad.s$^{-1}$, as seen on Fig. \ref{3d_cloak_sphere_elastic5}) of polarization $(1, 1, 1)$ located in a homogeneous isotropic medium  at a distance $8.062$~m from the origin. Real part of the deformation components $\varepsilon_{ij}=\frac{1}{2}(\frac{\partial u_i}{\partial x_j}+\frac{\partial u_j}{\partial x_i}),$ with $i,j=1,2,3,$ of the strain tensor $\varepsilon$.  Compare  with Figure \ref{3d_cloak_sphere_elastic3a} and Figure \ref{3d_cloak_sphere_elastic3b}.
}
\label{fig:strain_mixed_cloak_w2p20}
\end{figure*}

The mixed cloaks we introduce here consist of non-singular cloaks supplemented by an inner shell  made of  a Perfectly Matched Layer. These cloaks have the property to achieve both invisibility and protection (enhanced fields due to Mie type  resonances inside the soft sphere are drastically attenuated).  Figure  \ref{3d_cloak_sphere_elastic1} gives a description  of the geometric construction of the mixed cloak. The coefficients of the elasticity tensor and the density in the cloak and the PML  parts of the mixed cloak, are  the same as in  (\ref{sc}) and   (\ref{scpml}), respectively. Figures \ref{3d_cloak_sphere_elastic5}-(4), \ref{3d_cloak_sphere_elastic2_pml}, \ref{fig:strain_mixed_cloak} and \ref{fig:strain_mixed_cloak_w2p20} depict the corresponding numerical illustrations.  Compared to the case of non-mixed (ordinary) cloaks discussed in the previous sections, one realizes the absence of field inside the cloaked area, entailing a protection for the soft sphere. The results of a  comparative quantitative study between the different cases are  illustrated in Figure \ref{3d_cloak_sphere_elastic5}, where the evidence of a protection is seen. 
 However, it should be
noted that some reminiscence of the resonances  persists in Figures \ref{3d_cloak_sphere_elastic5} and  \ref{fig:strain_mixed_cloak_w2p20}, which might be attributed to trapped modes within the PML itself. 
As  they achieve both cloaking and protection throughout a large range of frequencies, mixed cloaks are thus foreseen to play an important role in the advent of real manufactured mechanical cloaks. A homogenization algorithm could be applied for both the PML and the cloak in order to seek existing materials for their practical realization.
 Such a concept of mixed cloaks is also valid  in the case of singular cloaks (in that case, one need simply add a PML layer at the singular boundary).  It can be extended to other waves areas in physics and engineering. 

\vspace{-5mm}
\section{Conclusion}
Finally, we would like to stress that main features of the spherical elastic cloaks
which we designed are their capability to make a soft sphere placed inside
invisible to incoming elastic waves, not from a wave protection viewpoint,
but from a substantial scattering reduction standpoint. Applications in anti-earthquake devices
would require some further analysis, insofar as the elastic field can be
dramatically enhanced within the cloak compared to a soft sphere on its own
(note in passing that we considered a simplified model for soft spheres with
real valued density and Lam\'e constants, although we appreciate that
these might be complex valued \cite{palmeri2005}).
However, if one has in mind to make a cloak for the range of frequencies of earthquakes
from $2$ to $5$ rad.s$^{-1}$, our design might form the elementary brick of a more elaborate
seismic cloak, which might be designed with a different geometric transform.
Moreover, the material parameters of our
cloaks are not frequency dependent, which is another pitfall since in practice one would fabricate some locally resonant structured materials in order to achieve the required density and elasticity tensor within the cloak, and we were
limited below $2$ rad.s$^{-1}$ by the accuracy of the spherical PML (the larger
the wave wavelength, the more absorption needed in PML) and above $5$ rad.s$^{-1}$
we are limited by the computational resources (numerical computation at $10$ rad.s$^{-1}$ required about 2 million tetrahedral elements for a converged result). A lot remains to be explored outside this range of angular frequencies.
However, our study opens a route to research in seismic metamaterials, which is a very immature (but fascinating) field \cite{kim2012,prl2014}. 
We finally note that recent advances in fabrication and characterization of elastic metamaterials \cite{kadic2012,kadic2013,kadic2014} could foster experiments in an approximate 3D elastic cloak. Of course, the metamaterial would only be able in
practice to display a strong anisotropy on the cloak's inner boundary and it would only work throughout a finite range of frequencies. Its properties could be derived for instance from an effective medium approach in a similar
way to what was proposed \cite{farhat2012} and experimentally validated \cite{stenger2012} for elastic waves in thin plates. Another route would be to use pre-stressed elastic
media which would naturally have the required elasticity tensor for cloaking to be fully operational over a broad range of frequencies \cite{william2012a,william2012b}.

As an alternative to elastodynamic cloaks with inner resonances, we further introduced the concept of mixed cloaks that have an additional PML layer attached to the inner boundary of the cloak. This PML considerably reduces the amplitude
of the trapped eigenstates within the invisibility region, without deteriorating the cloaking effect. In this way we achieved good protection against volume elastic wave over a broad range of frequencies. This concept of mixed cloaks could
be further translated into other wave areas for similar protection purpose.

The authors acknowledge European funding through ERC Starting Grant ANAMORPHISM.
The authors also wish to thank Dr. Younes Achaoui for insightful comments on wave propagation
in solid media.

\appendix\label{appendix1}

\section{Tensors of elasticity and deformation in spherical coordinates}
\vspace{-6mm}
Let us recall that any isotropic (possibly heterogeneous) elastic medium can be described by three (possibly spatially varying) scalar valued parameters $\lambda$, $\mu$ and $\rho$, which are respectively the Lam\'e parameters characterizing the compressibility and shear moduli, and the density, of the medium. The first two parameters are enough to generate the 21 non-zero components of the fully symmetric elastic constitutive tensor $\C$ written in a spherical basis (the reader can compare these expressions with those of the asymmetric elasticity tensor in Eq. \ref{sc}), for a homogeneous isotropic medium, which are  
\begin{eqnarray}
C_{rrrr}&=&
C_{\theta\theta\theta\theta}=C_{\phi \phi\phi\phi}=  (\lambda+2\mu),
\nonumber \\
C_{rr\theta\theta}&=&  C_{\theta\theta rr}=C_{rr\phi\phi}= C_{\phi \phi rr}
=C_{\theta\theta\phi\phi}=  C_{\phi\phi\theta\theta}=\lambda , ~\nonumber
\\
C_{r\theta r\theta}&=&  C_{r\phi r\phi}=
C_{\theta r\theta r}=  C_{\theta\phi \theta\phi}=C_{\phi r\phi r}=  C_{\phi\theta\phi \theta}
\nonumber
\\
&=&C_{r\theta\theta r}=  C_{\theta rr\theta}=  C_{r\phi\phi r}= C_{\phi rr \phi }
=C_{\theta\phi \phi\theta}\nonumber
=C_{\phi\theta\theta \phi}=\mu \;  . \;
\label{sc3}
\end{eqnarray}
\begin{widetext}
The gradient, of the displacement field  ${\bf u}=(u_r,u_\theta,u_\phi)$, has the following entries in spherical coordinates
\begin{equation}
 \nabla{\mathbf u}:=
 \begin{pmatrix}  \frac{\partial u_r}{\partial r}  &  \frac{\partial u_\theta}{\partial r} & \frac{\partial u_\phi}{\partial r} 
\\
 \frac{1}{r}\frac{\partial u_r}{\partial \theta} - \frac{u_\theta}{r} &   \frac{1}{r}\frac{\partial u_\theta}{\partial \theta} + \frac{u_r}{r}&\frac{1}{r} \frac{\partial u_\phi}{\partial \theta}
 \\
 \frac{1}{r\sin\theta}\frac{\partial u_r}{\partial \phi} - \frac{u_\phi}{r} &  \frac{1}{r\sin\theta}\frac{\partial u_\theta}{\partial \phi} -  \frac{\cos\theta}{r\sin\theta}u_ \phi&  \frac{u_r}{r} +  \frac{\cos\theta}{r\sin\theta}u_ \theta+ \frac{1}{r\sin\theta}\frac{\partial u_\phi}{\partial \phi}
 \end{pmatrix}.
\label{gradu}
\end{equation}
\end{widetext}

\section{Further numerics}
\begin{figure}[ht!]
\resizebox{85mm}{!}{\includegraphics{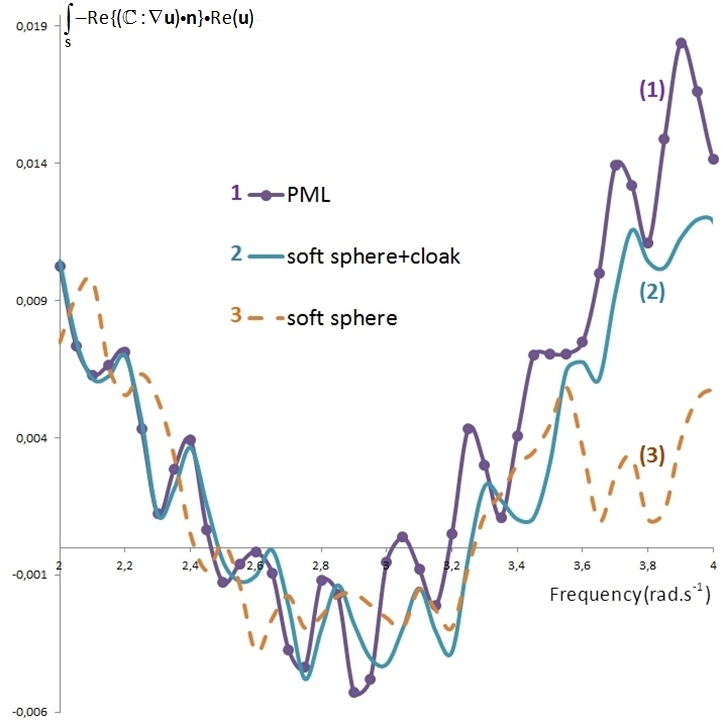}}
\vspace{-3mm}
\caption{The integral  $\displaystyle \int_{S}- Re \left[ \left(\C:\nabla {\bf u}\right)\cdot{\bf n}\right]\cdot Re({\bf u}) $ 
over a surface of sphere S of radius 7m, centered at the origin. A comparison between the isotropic homogeneous domain used as a benchmark (dotted curve (1)), the same domain now containing a soft sphere surrounded by a cloak (solid curve (2))  and  a soft sphere alone (dashed curve (3)). 
Note that the frequency varies by a step of 0.05 rad.s$^{-1}.$}
\label{flux_through_surfaces_alanorris}
\end{figure}

\begin{figure}[!htb]
\resizebox{85mm}{!}{\includegraphics{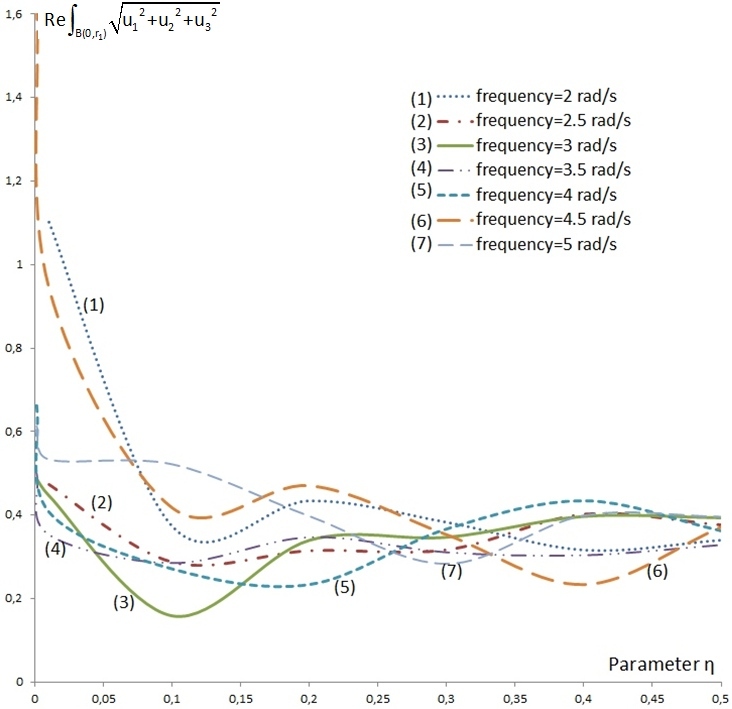}}
\vspace{-3mm}
\caption[!htb]{
Real part of the complex valued quantity $\int_{B({\bf 0},r_1)}\sqrt{u_1^2+u_2^2+u_3^2}$
of the displacement field inside the small isotropic homogeneous soft balls of radius $r_1=2~m$ surrounded by  cloak of outer radius
 $r_2=4~ m$, against the parameter $\eta$ varying from 0.001 to 0.5, 
for the same point force as in Fig. \ref{3d_cloak_sphere_elastic4}. 
Note the shift in frequencies with Fig. \ref{3d_cloak_sphere_elastic4}
where the L$^2$-norm of displacement field was computed.
}
\label{3d_cloak_sphere_elastic4appendix}
\end{figure}

\begin{figure}[!htb]
\resizebox{85mm}{!}{\includegraphics{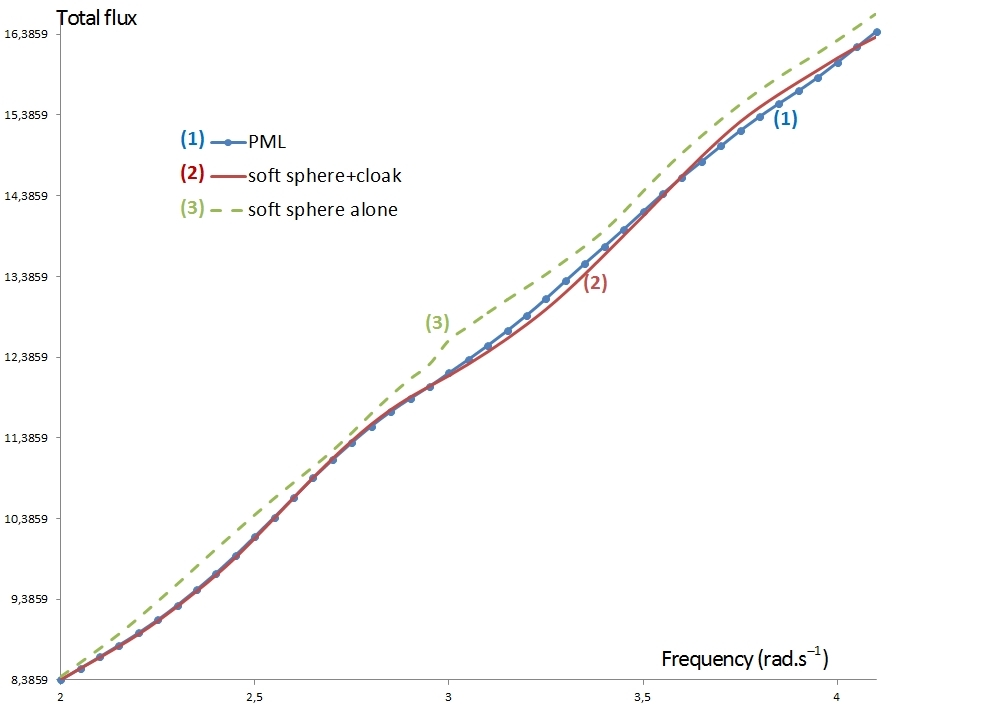}}
\vspace{-3mm}
\caption{Total flux $\int_S\mid\left[\left(\C:\nabla {\bf u}\right)\cdot{\bf n}\right]\cdot {\bf n}\mid$ 
on the surface of a sphere S of radius 7m, centered at the origin. A comparison between case (1) (solid line) of the isotropic homogeneous domain used as a benchmark, the same domain now containing a soft sphere with $\eta=0.3$ surrounded by a cloak (dotted line  (2))  and  then a soft sphere alone (dashed line (3)).  The point force is as in Fig. \ref{3d_cloak_sphere_elastic_mesh}. The free elastic space (1) and cloaked soft sphere (2) curves are nearly superimposed, whereas the soft sphere alone (2) is clearly above them. Note that the  frequency varies with a step of  0.05 rad.s$^{-1}.$}
\label{flux_through_surfaces}
\end{figure}

We now complement numerical analysis discussed within the main text with additional quantitative-- see Fig.\ref{flux_through_surfaces_alanorris}-\ref{flux_through_surfaces}-- evidence of cloaking.
One notes that in Fig.\ref{flux_through_surfaces_alanorris} cloaking seems to degrade for frequencies higher than $2.8$ rad.s$^{-1}$, which could be attributed to the fact that the mesh is no longer fine enough to ensure fully accurate numerical results (we have made some numerical tests to double-check this origin of inaccuracy).   

\medskip

Using a measure of flux of energy similar to that of Norris in Eq.(4.1) of \cite{norris-scat} which is based upon the total field instead of the scattered field alone, we provide yet another numerical proof of suppression of scattering by the soft elastic sphere when it is surrounded by the cloak, see Fig. \ref{flux_through_surfaces_alanorris}.
We note that the dotted and solid curves, corresponding to elastic free space (1) and cloaked soft sphere (2) are nearly superimposed between $2$ and $2.5$ rad.s$^{-1}$, which tells us that numerics are fully accurate in this frequency range. However, for frequencies between $2.5$ and $4$ rad.s$^{-1}$ curves (1) and (2) can be clearly
told apart, an effect which could be attributed to the fact that the mesh is perhaps too loose to ensure full numerical convergence of the solution (we kept the same mesh for all frequencies). Nevertheless, the dashed curve (3) that corresponds to scattering by a soft sphere is always the curve farther apart from (1). Let us stress again that we have achieved this with spherical cloaks having elasticity tensors without the minor symmetries, akin to Cosserat media \cite{cosserat}. 


.
\newline
\newline
\newline
\newline
\newline
\newline

\end{document}